\documentclass[conference]{IEEEtran}
\usepackage{cite}

\ifCLASSINFOpdf
\usepackage[pdftex]{graphicx}
\else
\fi
\usepackage{epstopdf}
\usepackage[T1]{fontenc}

\usepackage{amsmath}
\usepackage{algorithmic}

\usepackage{array}

\ifCLASSOPTIONcompsoc
  \usepackage[caption=false,font=normalsize,labelfont=sf,textfont=sf]{subfig}
\else
  \usepackage[caption=false,font=footnotesize]{subfig}
\fi
\usepackage{url}

\begin{document}
%
\title{Localised response retrieval from Hamamatsu H9500 for a coded aperture neutron-gamma imaging system based on an organic pixelated plastic scintillator (EJ-299-34)}

\author{\IEEEauthorblockN{Michal J. Cieslak}
\IEEEauthorblockA{Engineering Department\\Lancaster University\\
Lancaster, LA1 4YW\\
United Kingdom\\
Email: m.cieslak@lancaster.ac.uk}
\and
\IEEEauthorblockN{Kelum A.A. Gamage}
\IEEEauthorblockA{School of Engineering\\
University of Glasgow\\
Glasgow, G12 8QQ\\
United Kingdom\\
Email: kelum.gamage@glasgow.ac.uk}}


%


\maketitle

\begin{abstract}
Localised response of a sensitive light detector, such as Hamamatsu H9500 multi anode photomultiplier tube (MAPMT), is of vital importance for coded-aperture imaging systems. When coupled with a suitable sensitive detector (e.g. EJ-299-34 plastic scintillator), output signals of the MAPMT can be retrieved to infer the interaction location in the pixelated scintillator. Given the number of anodes in H9500 (256), significant processing power would be necessary to handle each pulse individually. Therefore, a readout electronics system was developed, based on resistive network approach, which reduces the number of output signals to individual X and Y coordinates, and subsequently allows particle identification. Coordinates retrieved in this manner can be analysed in real time and used to infer the two-dimensional location. Particle type can be also exploited by pulse shape discrimination (PSD) application to the scintillator's response. In this study, 169 anodes were used (due to coded-aperture design rules), and reduced to two X and Y output signals. These have been digitised using a bespoke FPGA based two channel 14-bit 150 MSPS digitiser. The digital data are transferred to a host application using UART to USB converter operating at 12 Mbits/s. Promising results have been observed when the scintillator's response was tested in single particle field of $^{137}$Cs. However, further tests performed in the mixed-field environment of $^{252}$Cf suggest that faster digitiser may be required to obtain the required PSD performance.       
\end{abstract}


%
\IEEEpeerreviewmaketitle

\section{Introduction}
Neutron radiation is an example of highly penetrating radiation, which if not controlled, can pose a significant threat from nuclear safety and security viewpoint. Therefore, effective neutron detection it is of vital importance in areas, such as medicine \cite{Clarke2016}, border control \cite{Kouzes2008} and nuclear decommissioning \cite{Jallu2012}. Moreover, the localisation of neutron emitting isotopes can be particularly challenging in nuclear decommissioning applications, where large areas may be contaminated and exact location of the radioactive substance unknown.  Collimator based imaging systems with a single opening in the aperture, can be used to scan the large areas and create an image identifying the location of the neutron emitters \cite{Gamage2012}. However, the drawbacks of such system include the fact that process of image creation can be time consuming (due to scanning requirement) and the effective field-of-view much limited.

\par

Coded-aperture neutron imaging systems show promising results when considered as an alternative to the single opening collimators. Relatively large scale neutron detectors utilising coded-aperture approach have been previously tested and characterised \cite{Hausladen2013}, \cite{Griffith2017}. Both examples exploit pulse shape discrimination (PSD) capabilities of organic liquid (EJ-309) and plastic (EJ-299) scintillators. Small scale implementation of the coded-aperture imaging was previously realised with CLYC (Cs2LiYCl6:Ce) inorganic scintillation block as the radiation sensitive detector \cite{Whitney2014}. 

\par 

In this study a pixelated organic plastic scintillator (Fig. 1) was coupled to Hamamatsu H9500 MAPMT. The size of the single pixel has been matched to the size of a single anode of Hamamatsu H9500 (2.8mm x 2.8 mm) to reduce the risk of crosstalk between the anodes. Also, each scintillator block (2.8 mm x 2.8 mm x 15 mm) has been covered with 3M reflective tape to avoid interference between adjoining blocks. Given the sensitivity of the scintillator to both neutrons and gamma-ray photons, such configuration requires a bespoke readout electronics system. It is required that the readout electronics can not only infer the location of the interaction, but also enable particle classification (either neutron or gamma-ray photon).

\begin{figure}[h]
\begin{center}
\includegraphics[width = 6cm, height = 4cm]{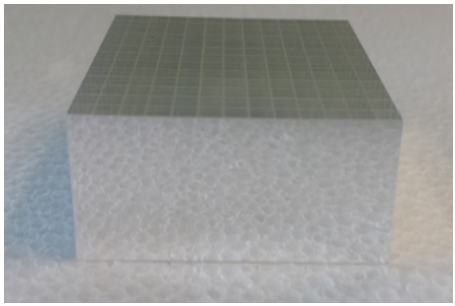}
\caption{Solid pixelated organic plastic scintillator (EJ-299-34).}
\label{fig1}
\end{center}
\end{figure}

\section{Methodology}

The readout electronics system, consists of resistive network and pulse shaping circuit followed by Field Programmable Gate Array (FPGA) based signal digitiser. The resistive network, first proposed by Popov \cite{Popov2011}, splits the voltage incurred as a result of an interaction in the specific pixel into equal parts on X and Y axis. This is then passed on to the weighted pre-amp pulse shaping stage, where based on the voltage detected, X and Y coordinates of the interaction can be computed. 

\par 
This stage has been simulated using TI-TINA (SPICE based) simulation package. Simulated neutron pulses were applied to three adjoining anodes and transient response recorded at the relative outputs of the pre-amp stage. Fig. 2 presents the proportional response of the simulated circuit at selected outputs, which advocates the claim that two-dimensional coordinates of the interaction can be inferred using this method. Furthermore, the resistive network does not appear to distort the shape of the applied pulse, presenting the potential of PSD techniques application for particle identification.

\par

Following the successful simulation of the circuit, resistive network was designed and built into a printed circuit board (PCB). It contains 507 surface-mount resistors in 803 and 605 packages, split onto four printed circuit boards (PCBs), which are directly attached to the output connectors of the Hamamatsu H9500 MAPMT. Complete assembly attached to the MAPMT is presented in Fig. 3. It reduces the number of output signals from 169 to 26. This is further reduced to 2 signals (X and Y coordinates) through the shaping pre-amp and summing op-amp circuitry.

\begin{figure}[h]
\begin{center}
\includegraphics[width = 9cm, height = 6cm]{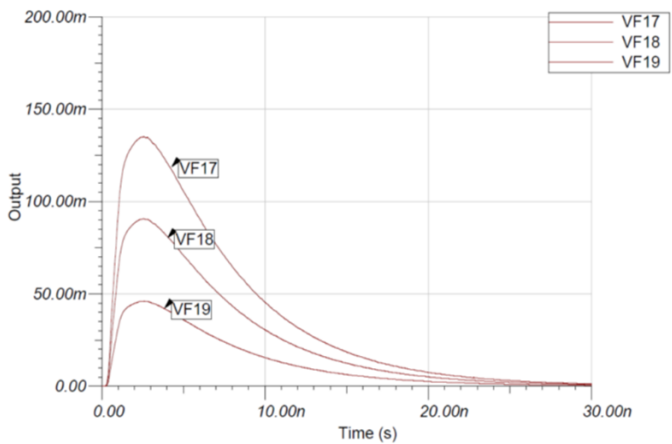}
\caption{Proportional response of the circuit simulated in TI-TINA software package for pulses connected to three adjoining anodes on one axis. VF17-VF19 correspond to transient response to the simulated neutron pulse connected to the three adjoining anodes.}
\label{fig2}
\end{center}
\end{figure}

\begin{figure}[h]
\begin{center}
\includegraphics[width = 8cm, height = 4cm]{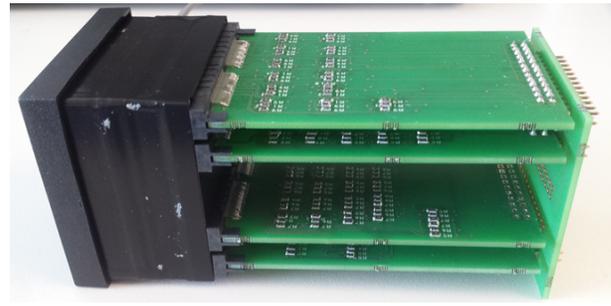}
\caption{A network of 507 resistors that reduces the number of output signals from 169 to 26 (13 on X and 13 in Y axis).}
\label{fig3}
\end{center}
\end{figure}

\par

Analogue pulses present at the output stage of the summing op-amp circuitry are digitised utilising two Analog Devices AD9254 150 MSps, 14-bit amplitude resolution ADCs directly linked to Altera Cyclone IV EP4CE115 FPGA. Verilog VHDL code has been developed that reads the outputs of the two 14-bit ADCs and transfers the data at 12 Mbits/s to a host application (such as Python application on a Laptop running Linux) via USB to UART converter. Each sample is transmitted to the host application with a channel identifier (A or B), e.g. A4087, B6743. 

\par

In order to further assess PSD capability of the pixelated scintillator, the anode of the  PMT was directly linked to a 10-bit, 500 MSps "raw" data digitizer where the data were transferred from the digitizer to a host PC using Ethernet protocol. The drawback of this system is single channel operation. However, much higher sampling rate is expected to allow more accurate pulse reconstruction despite the reduced ADC resolution.  

\section{Results and Conclusions}

Simulation of the resistive network deem a result (Fig. 2) that is proportional to the location of the interaction of the incoming particle, suggesting localised response of the MAPMT can be obtained in this way. Additionally, pulse shape analysis of the response of the pixelated scintillator to single particle field of $^{137}$Cs proposes PSD as a viable technique for particle identification, given the comparison to modelled pulses, as presented in Fig. 4.

\begin{figure}[h]
\centering
\subfloat[]{\includegraphics[width=8cm]{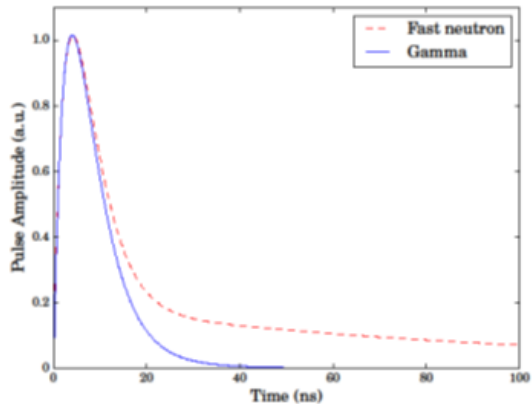}
\label{fig4a}}
\hfil
\subfloat[]{\includegraphics[width=8cm]{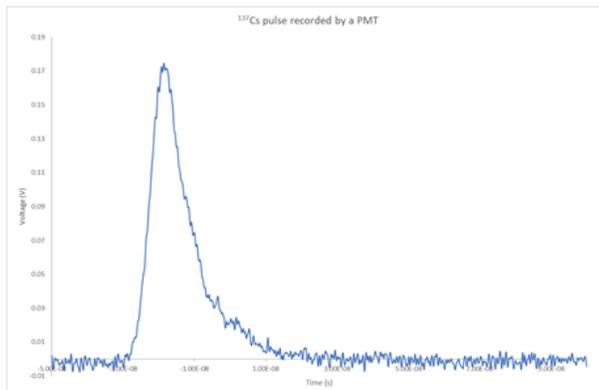}
\label{fig4b}}
\caption{ A comparison of a) A mathematical model of gamma-ray photon and neutron induced pulses from organic scintillator, b) A recoded gamma-ray pulse from $^{137}$Cs.}
\label{fig4}
\end{figure}

\par

Initial simulation and experimental results obtained, based on the single particle type detection, present a potential for good localisation and particle separation capabilities of the complete system. Given the neutron-gamma separation potential  of the scintillator, the complete system was also tested in dual-particle environment of the spontaneous fission field of $^{252}$Cf. 

\par

Two different digitizer systems were tested, as initial results obtained using the FPGA based 150 MSps digitizer system failed to discriminate neutrons from gamma-ray photons. Results obtained from the 500 MSps "raw" data digitizer (Fig. 5) show somewhat promising PSD potential, given the modulated energy spectrum of $^{252}$Cf at Lancaster University. However, the estimated figure-of-merit (FOM) of 0.337 suggests considerably poorer performance when compared with the FOM of 0.649 previously quoted in the literature for the cylindrical plastic scintillator samples tested in the same environment \cite{Cieslak2017}. Additional tests and analysis with "bare" mixed-field sources would also be advantageous to further assess the suitability of the system for similar applications. 

\begin{figure}[!h]
\begin{center}
\includegraphics[width = 8cm, height = 6cm]{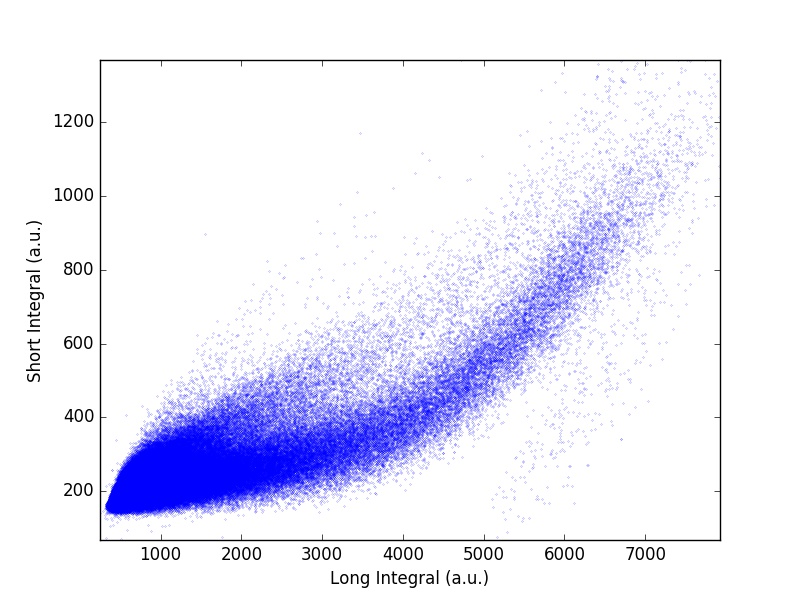}
\caption{PSD scatter plots using CCM for Pixelated EJ-299-34 plastic scintillator).}
\label{fig5}
\end{center}
\end{figure}

\section*{Acknowledgment}

The authors would like to acknowledge the funding support from EPSRC via Faculty of Science and Technology, Lancaster University, U.K. and Sellafield Ltd.

\end{document}